\documentclass[twocolumn,showpacs,preprintnumbers,amsmath,amssymb,APSl,prd,nofootinbib,superscriptaddress]{revtex4-1}

\usepackage{dcolumn}
\usepackage{bm}
\usepackage{ifpdf}
\usepackage{hyperref}
\usepackage{bm}
\usepackage{xcolor,color,graphicx,graphics}
\usepackage[spanish,english]{babel}
\usepackage[latin1]{inputenc}
\usepackage[OT1]{fontenc}
\usepackage{latexsym,amssymb,amsmath,amsfonts}
\usepackage{makeidx}
\usepackage{epsfig,subfigure}
\usepackage{natbib}
\usepackage{epstopdf}
\usepackage{mathrsfs}
\usepackage{hyperref}
\hypersetup{colorlinks=true, linkcolor=blue, citecolor=green}
\usepackage{enumerate}

\definecolor{red}{rgb}{1,0,0}

\def\+{^\dagger}

\def\<{\leftarrow}
\def\>{\rightarrow}

\def\({\left(}
\def\){\right)}

\def\t{\tau}

\def\sech{\mathop{\rm sech}\nolimits} 

\newcommand{\bi}{\begin{itemize}} 				\newcommand{\ei}{\end{itemize}}
\newcommand{\benu}{\begin{enumerate}} 		\newcommand{\enu}{\end{enumerate}}
\newcommand{\bd}{\begin{dinglist}{0}}     \newcommand{\ed}{\end{dinglist}}
\newcommand{\bfig}{\begin{figure}[htbp]}  \newcommand{\efig}{\end{figure}}
        			
\newcommand{\bc}{\begin{center}} 				  \newcommand{\ec}{\end{center}}
\newcommand{\be}{\begin{equation}} 				\newcommand{\ee}{\end{equation}}
\newcommand{\bsub}{\begin{subequations}}  \newcommand{\esub}{\end{subequations}}
\newcommand{\ben}{\begin{eqnarray}} 			\newcommand{\een}{\end{eqnarray}}
\newcommand{\ba}[1]{\begin{array}{#1}} 		\newcommand{\ea}{\end{array}}
\newcommand{\bea}{\begin{equation}\begin{array}{rcl}}
\newcommand{\eea}{\end{array}\end{equation}}

\begin{document}
\title{Nonsingular black holes in nonlinear gravity  coupled to Euler-Heisenberg electrodynamics}

\author{Merce Guerrero} \email{merguerr@ucm.es}
\author{Diego Rubiera-Garcia} \email{drubiera@ucm.es}
\affiliation{Departamento de F\'isica Te\'orica and IPARCOS,
	Universidad Complutense de Madrid, E-28040 Madrid, Spain}

\date{\today}
\begin{abstract}
We study static, spherically symmetric black holes supported by Euler-Heisenberg theory of electrodynamics and coupled to two different modified theories of gravity. Such theories are the quadratic $f(R)$ model and Eddington-inspired Born-Infeld gravity, both formulated in metric-affine spaces, where metric and affine connection are independent fields. We find exact solutions of the corresponding field equations in both cases, characterized by mass, charge, the Euler-Heisenberg coupling parameter and the modified gravity one. For each such family of solutions, we characterize its horizon structure and the modifications in the innermost region, finding that some subclasses are geodesically complete. The singularity regularization is achieved under two different mechanisms: either the boundary of the manifold is pushed to an infinite affine distance, not being able to be reached in finite time by any geodesic, or the presence of a wormhole structure allows for the smooth extension of all geodesics overcoming the maximum of the potential barrier.
\end{abstract}

\maketitle

\section{Introduction} \label{sec:I}

Black holes are one of the most fascinating objects in Nature. Originally obtained as exact solutions of Einstein's field equations, their properties were poorly understood for decades until becoming nowadays a full-fledged member of the family of astronomical objects. From a mathematical point of view, they can be formed from a regular distribution of matter  in such a way that a trapped surface is developed \cite{JoshiBook}. From an astrophysical viewpoint, the gravitational collapse out of fuel-exhausted main-sequence stars ($\gtrsim 25M_{\odot}$) provides the physical mechanism for such a  generation \cite{Heger:2002by}. Moreover, no matter the properties and/or symmetries of the original configuration, the outside metric to the end-state of such a collapse will be always the Kerr-Newman solution, described solely by three parameters: mass, charge and angular momentum \cite{SteBook}. Over the years, we have accumulated plenty of evidence on the reliability of the simpler Kerr solution (since charge can be typically neglected in astrophysical environments \cite{Zajacek:2019kla}) to describe such objects, as follows from observations of the X-ray radiation emitted from the inner part of their accretion disks \cite{Bambi:2017iyh,Bambi:2019xzp}, from gravitational wave emission out of binary mergers \cite{Abbott1,Abbott2} and from the imaging of the neighborhood ({\it i.e.} the shadow) of the central supermassive object of the M87 galaxy \cite{M87}.

Given the remarkable agreement between the theoretical predictions and the astronomical observations regarding black holes, it is worth taking seriously another key feature of their theoretical description, namely, the existence of space-time singularities deep inside them \cite{WaldBook}. Indeed, as long as General Relativity (GR) holds, the theorems on singularities \cite{Theorem1,Theorem2} (for a pedagogical discussion see e.g. \cite{Senovilla:2014gza}) tell us that the development of a focusing point at a finite affine time for some set of geodesics in the innermost region of a black hole ($r=0$ in the case of a spherically symmetric black hole)  is unavoidable provided that standard energy conditions are satisfied by the matter fields. Now, since null and time-like geodesics are associated to the trajectories of light rays and the free-falling of physical observers, respectively, the incompleteness of any of them is an utterly unpleasant feature, being linked to the lack of predictability of our physical theories. Typically, cosmic censorship arguments are developed \cite{CCC} in order to cover such singularities behind an event horizon, so as not to have observable effects on asymptotic observers. However, it is distressing that one needs to hide under the carpet such an abhorrent feature of an otherwise observationally successful object (outside its event horizon). Therefore, several arguments have been developed in order to overcome this difficulty without jeopardizing the exterior physics to the horizon, the jewel of the crown being the hyphotesis that quantum gravity effects should come to rescue when the growth of curvature approaches the Planck scale \cite{OritiBook}.

The question on how to incorporate such effects has received many different answers along the decades. To play as conservatively as possible, one way to address them is via effective modifications of the gravitational action \cite{Review1,Review2,Nojiri:2017ncd}, which could be able to provide some hints on the transition from the classical (GR) regime to the quantum non-classical one and, moreover, to yield new phenomenology in astrophysical environments \cite{Berti}. This way one can still safely use the tools of the differential manifolds paradigm but modifying/reinterpreting some of its building blocks. In the present work we shall follow this path, and  adhere to the spirit of the Equivalence Principle, guaranteeing the universality of free-fall motion while keeping the minimal coupling of the matter fields to the gravitational sector but introducing three major modifications to the usual GR dynamics. The first one is simply to restore the metric and the affine connection to their status as independent entities (Palatini or metric-affine approach \cite{Olmo:2011uz}). Indeed, GR can be consistently formulated as a metric-affine theory, with the variation of the Einstein-Hilbert action with respect to the independent connection yielding the metric-connection compatibility condition. As a consequence, the predictions of this formulation are exactly the same ones as those of metric-formulated GR, where the affine connection is imposed {\it ab initio} to be given the Christoffel symbols of the metric \cite{BeltranJimenez:2019tjy}. 

Since the metric-affine formulation of GR does not introduce any new predictions, the second major modification is to consider more general actions, where such a dynamics of metric-affine gravities strongly departs from their metric cousins, offering new ways of addressing the issue with space-time singularities. For the sake of this paper, we shall consider two well known gravitational extensions of GR: quadratic $f(R)$ gravity \cite{Staro} and Eddington-inspired Born-Infeld (EiBI) gravity \cite{EiBI}. The underlying reason for this choice is that they belong to a more general class of theories dubbed as  Ricci-Based Gravities (RBGs), which are those built with scalars from contractions of the metric with the (symmetric part of the) Ricci tensor. These theories yield second-order, ghost-free equations compatible with all solar system tests and gravitational wave observations carried out so far.  Moreover, the new metric-affine dynamics is triggered in different ways: while in the $f(R)$ case it is oblivious to anything but to the trace of the stress-energy tensor, in the EiBI case it has access to the full stress-energy tensor. 

The third major modification lies in the matter sector, where we shall use an old acquaintance of the singularity-regularization attempts: nonlinear electrodynamics \cite{AyonBeato:1998ub,Bronnikov:2000vy,Dymnikova:2004zc,Balart:2014cga,Balart:2014jia,Dymnikova:2015hka,Rodrigues:2015ayd,Nojiri:2017kex,Chinaglia:2017uqd,Rodrigues:2018bdc,Rodrigues:2019xrc,Allahyari:2019jqz}. Indeed, given the tracelessness of the stress-energy tensor of Maxwell electrodynamics, which would therefore yield the same solutions for $f(R)$ theories as the GR ones, this pick will allow us to compare the predictions of $f(R)$ and EiBI gravity on an equal footing. On more physical grounds, it is known that way before the scale where quantum gravity effects are expected to be excited in the innermost region of black holes, the growth of the electric field would induce quantum vacuum polarization effects modifying the classical description of Maxwell electrodynamics. In an effective approach, such effects to one loop and in the slowly-varying approximation can be incorporated by adding a quadratic piece in the electromagnetic field invariants to the Maxwell Lagrangian, yielding the so-called Euler-Heisenberg electrodynamics \cite{EH,EffBook}. Within GR such a model has some nice properties, like a finite energy associated to the system of point-like charges and the existence of new gravitational configurations in terms of the structure of horizons \cite{Yajima:2000kw,Ruffini:2013hia,Breton:2019arv}. However, space-time singularities still plague all such configurations, and similar comments apply to any nonlinear electrodynamics satisfying physically reasonable conditions \cite{Bronnikov:2000yz}. 

The main aim of this work is thus to find static, spherically symmetric solutions corresponding to quadratic $f(R)$ and EiBI gravity coupled to Euler-Heisenberg electrodynamics and investigate the existence of nonsingular black holes in both frameworks. We shall find that both of them have a branch of solutions (as given by the combination of the sign of the gravity and matter parameters) allowing for the completeness of all null and time-like geodesics. This restoration of geodesic completeness is achieved via two different mechanisms:  in the first one the focusing point is pushed out to an infinite affine distance preventing any set of geodesics to reach it in finite affine time, while in the second one a defocusing sphere is created at some finite affine distance represented by a wormhole throat with a finite area, in such a way that those geodesics able to overcome the potential barrier can be smoothly extended through the throat to another asymptotically flat region of the manifold. Rather than making quantitative predictions based on the scales where gravity/matter corrections should presumably appear, our aim is to qualitatively discuss the singularity-avoidance resolution mechanisms within these theories and how they fit within general studies aimed to achieve singularity-avoidance without breaking basic mathematical requirements or getting into contradiction with observations. 

This work is organized as follows: in Sec. \ref{sec:II} we introduce the basic framework in terms of metric-affine gravities and Euler-Heisenberg electrodynamics. In Sec. \ref{sec:III} we find spherically symmetric solutions for quadratic $f(R)$ gravity and discuss their properties, with particular emphasis on the horizon and geodesic structure. A similar analysis is carried out for EiBI gravity in Sec. \ref{sec:IV}. Finally in Sec. \ref{sec:V} we summarize our findings and further discuss our results.

\section{Theoretical framework} \label{sec:II}

\subsection{Ricci-based gravities}

For the sake of this work, we shall establish the theoretical framework for the subclass of metric-affine gravities dubbed as Ricci-based gravities (RBGs), defined by the action
\begin{equation} \label{eq:action}
\mathcal{S}_m=\frac{1}{2\kappa^2} \int d^4x \sqrt{-g} \mathcal{L}_G(g_{\mu\nu},R_{\mu\nu}(\Gamma)) + \mathcal{S}_m(g_{\mu\nu},\psi_m) \ ,
\end{equation}
where $\kappa^2$ is Newton's constant in suitable units, $g$ is the determinant of the space-time metric $g_{\mu\nu}$, the Ricci tensor (assumed to be symmetric) $R_{\mu\nu}(\Gamma) \equiv {R^\alpha}_{\mu\alpha\nu}(\Gamma)$ is solely built out of the (torsion-free \cite{Afonso:2017bxr}) affine connection  $\Gamma \equiv \Gamma_{\mu\nu}^{\lambda}$ and the matter action $\mathcal{S}_m=\int d^4x \sqrt{-g} \mathcal{L}_m(g_{\mu\nu},\psi_m)$ is assumed to depend only on the space-time metric and on a set of matter fields $\psi_m$ but not on the connection, to ensure the fulfillment of the equivalence principle. These constraints upon the building blocks of the action (\ref{eq:action}) guarantee the second-order and ghost-free character of their field equations \cite{BeltranJimenez:2019acz,Jimenez:2020dpn}. Moreover, they imply that RBGs do not propagate extra degrees of freedom beyond the two polarizations of the gravitational field of GR and may pass solar system tests provided that the modifications to GR occur in the ultraviolet limit.

From independent variation of the action (\ref{eq:action}) with respect to metric and connection, the resulting field equations may be conveniently written in the Einstein-like representation
\begin{equation} \label{eq:eomRBG}
{G^\mu}_{\nu}(q)=\frac{\kappa^2}{\vert \Omega \vert^{1/2}} \left[{T^\mu}_{\nu}-\delta^\mu_\nu \left(\mathcal{L}_G + \frac{T}{2} \right)\right] \ ,
\end{equation}
where ${G^\mu}_{\nu}(q)$ is the Einstein tensor of a new rank-two tensor $q_{\mu\nu}$ satisfying $\nabla_{\mu}(\sqrt{-q} q^{\alpha\beta})=0$ (so that $\Gamma$ is Levi-Civita of $q$). This tensor is related to the space-time metric as
\begin{equation} \label{eq:Omegadef}
q_{\mu\nu}=g_{\mu\alpha}{\Omega^\alpha}_{\nu} \ ,
\end{equation}
where the explicit shape of the \emph{deformation matrix}  ${\Omega^\mu}_{\nu}$ (vertical bars denoting its determinant) depends on the particular $\mathcal{L}_G$ chosen, but can be written always (and so does $\mathcal{L}_G$ itself) on-shell in terms of the stress-energy tensor $T_{\mu\nu} \equiv \frac{2}{\sqrt{-g}}\frac{\delta \mathcal{S}_m}{\delta g^{\mu\nu}}$ (with $T$ denoting its trace) and possibly the space-time metric as well. Therefore, the right-hand side of the field equations (\ref{eq:eomRBG}) can be read off as representing an effective stress-energy tensor \cite{Afonso:2018bpv}.

\subsection{Euler-Heisenberg electrodynamics}

Nonlinear electrodynamics (NED) are described by a Lagrangian density
\begin{equation}
\mathcal{L}_m=\varphi(X,Y) \ ,
\end{equation}
where $ X= -\frac{1}{2} F_{	\mu \nu}F^{\mu \nu} $ and $Y=-\frac{1}{2}F_{\mu\nu}F^{*\mu\nu}$ are the two electromagnetic field invariants which can be built out of the field strength tensor, $ F_{\mu \nu}= \partial_\mu A_\nu - \partial_\nu A_\mu $, and its dual $F^{*\mu\nu}=\frac{1}{2}\epsilon^{\mu\nu\alpha\beta}F_{\alpha\beta}$. The corresponding field equations are written as $\nabla_{\mu}(\varphi_X F^{\mu\nu} + \varphi_Y F^{* \mu\nu})=0$, where $\varphi_X \equiv \frac{\partial \varphi}{\partial X}$ and $\varphi_Y \equiv \frac{\partial \varphi}{\partial Y}$. For electrostatic configurations, the only non-zero component  is $  F_{tr} \equiv E(r)$, and the field equations in {\it any} static, spherically symmetric space-time, $ds^2=g_{tt}^2+g_{rr}^2dr^2+r^2(d\theta^2 + \sin^2\theta d\phi^2)$, can be written as
\begin{equation} \label{eq:NEDeom}
X\varphi_X^2=\frac{q^2}{r^4} \ ,
\end{equation}
where $X \equiv E^2$ and $q$ is an integration constant identified as the electric charge for a given configuration. The NED stress-energy tensor
\begin{equation}
{T^\mu}_{\nu}=2(\varphi_X {F^\mu}_{\alpha}{F^\alpha}_{\nu} -\varphi_Y {F^\mu}_{\alpha}{F^{*\alpha}}_{\nu} )-\varphi(X,Y) \ ,
\end{equation}
for electrostatic configurations can be conveniently split into $2 \times 2$ blocks as
\begin{eqnarray}\label{stress-energy tensor}
{T^\mu}_\nu &=& \dfrac{1}{8\pi}\left(\begin{array}{cc} (\varphi-2X\varphi_X)\, \hat{I}_{2\times 2} &  \hat{0\,}_{2\times 2}\\  \hat{0\,}_{2\times 2}& \varphi \, \hat{I}_{2\times 2}\end{array}\right) \ ,
\end{eqnarray}
where $ \hat{0}_{2\times 2} $ and $ \hat{I}_{2\times 2} $ are the $2 \times 2$ zero and identity matrices, respectively. From this expression, the trace reads $T=\frac{1}{2\pi}(\varphi-X\varphi_X)$, which is non-vanishing as long as $\varphi \neq X$ (Maxwell electrodynamics).

For the sake of this paper, we shall restrict our considerations to the case of Euler-Heisenberg electrodynamics, which is described by the particular function\footnote{When considering the effective limit of quantum electrodynamics this Lagrangian picks another term in $Y$, which is vanishing for the electrostatic configurations of this paper. In such a limit, $\beta$ takes the value $\beta=\frac{2\alpha^2}{45m_e^2}$ \cite{Schwinger}, where $m_e$ is electron's mass and $\alpha$ the fine structure constant. However, for the purposes of this paper we shall take $\beta$ as a free parameter assuming only $\beta>0$.}
\begin{equation}\label{Eq:Euler-Heisenberg}
\varphi(X) = X\, + \, \beta X^2 \ .
\end{equation}
For this theory, the NED field equations (\ref{eq:NEDeom}) for (electro-) static, spherically symmetric fields read
\begin{equation} \label{eq:EEH}
E+2\beta E^3=\frac{q}{r^2} \ .
\end{equation}
As can be easily verified from this equation, at $r \rightarrow \infty$ this theory recovers the Coulomb field, $E=q/r^2$, while for $r=0$ we have instead $E=(q/2\beta)^{1/3} r^{-2/3}$.

The field equations (\ref{eq:NEDeom}) for EH electrodynamics can be solved in exact form as
\begin{equation}\label{Eq:field_invariant}
X(r)= \left[ \sqrt[3]{U  + \sqrt{V^3 + U^2}} \,+ \, \sqrt[3]{U  - \sqrt{V^3 + U^2}}\;  \right]^2 \ ,
\end{equation}
where $ U = \dfrac{q}{4 \beta r^2} $ and $ V = \dfrac{1}{6 \beta} $. To work with dimensionless variables, let us introduce a new length scale as
\begin{equation} \label{eq:rcdef}
r_{c}^4 =54\pi  l^2_\beta \, r_q^2 \ ,
\end{equation}
with $ l^2_\beta= \beta/\kappa^2 $ the squared NED length and $ r^2_q = \kappa^2 \, q^2/(4\pi) $ the squared charge radius. This way, Eq.\eqref{Eq:field_invariant} can be rewritten in terms of the dimensionless coordinate $ z=r/r_c $ as
\begin{equation}\label{Eq:X_red}
X(z)=\dfrac{1}{6 \beta\, z^{4/3}}\left[ \left( 1+\sqrt{1+z^4}\right)^{1/3}+\left( 1-\sqrt{1+z^4}\right)^{1/3} \right]^2.
\end{equation}
Moreover, we can get rid of cubic roots via the alternative expression \cite{Kruglov:2017ymn}
\begin{equation} \label{eq:Xz}
X(z)=\frac{2}{3\beta} \text{Sinh}^2\left[\frac{1}{3} \ln \left(\frac{1}{z^2}\left(1+ \sqrt{z^4+1}\right)\right)\right].
\end{equation}
In terms of this dimensionless variable the stress-energy tensor (\ref{stress-energy tensor}) for EH electrodynamics reads
\begin{eqnarray}\label{final-stress-energy tensor}
{T^\mu}_\nu &=& \dfrac{1}{8\pi} \left(\begin{array}{cc} -X (1+3\beta X)\, \hat{I}_{2\times 2} &  \hat{0\,}_{2\times 2}\\  \hat{0\,}_{2\times 2}& X (1+\beta X) \, \hat{I}_{2\times 2}\end{array}\right)
\end{eqnarray}
such that its components are found upon substitution of (\ref{eq:Xz}). Note that, upon replacement of Eq.(\ref{eq:EEH}) in the ${^t}_t$ component of the stress-energy tensor (\ref{final-stress-energy tensor}) one finds that in the $r \to 0$ limit, the contribution to the energy of the EH field behaves as $ \sim  \int {T^0_0} r^2 dr \sim r^{1/3} \to 0$, which implies that the total energy associated to electrostatic configurations in EH electrodynamics is finite. On the other hand, from this form of the stress-energy tensor it can be verified that EH electrodynamics satisfies the weak energy condition provided that $\beta>0$, which is the case studied in this work. Moreover, when coupled to the Einstein-Hilbert action of GR, the finite character of the electrostatic solutions of this theory manifests in the fact that, besides configurations with two or a single (degenerate) horizons and naked singularities, typical of the Reissner-Nordstr\"om solution of GR, there are also configurations with a single non-degenerate horizon (resembling the Schwarzschild black hole). However, in all cases, a singular behaviour is found as follows from the geodesic incompleteness of all such solutions (see however \cite{Poshteh:2020sgp}).

\section{Quadratic $f(R)$ Gravity} \label{sec:III}

\subsection{Derivation of the solution}

Our first RBG model to be analyzed is quadratic  $ f(R) $ gravity, given by the Lagrangian density
\begin{equation}\label{f(R)}
\mathcal{L}_G= f(R) = R + \alpha R^2 \ ,
\end{equation}
where  $ \alpha $ is a constant with dimensions of length squared\footnote{In an approach of quantization of fields in curved space-times the Einstein-Hilbert action of GR should pick, on the ultraviolet limit, additional curvature corrections suppressed by powers of Planck's length squared $l_P^2$ \cite{PTBook,BDBook}. Therefore, one would be tempted to interpret $\alpha$ as $l_P^2$ times a constant of size unity, and the same would apply to EiBI parameter $\epsilon$ in Sec. \ref{sec:III}. However, we shall refrain ourselves to make such identifications and take $\alpha$ as a free parameter for the sake of this work.}.
For $f(R)$ gravity, the trace of the RBG field equations provides us with the following relation $ R\, f_R -2f = \kappa^2 T $ (with $f_R \equiv df/dR$), which tells us that the curvature scalar can be removed in favour of the trace of the stress-energy tensor. This fact implies that only NEDs with a non-vanishing trace will yield new dynamics as compared to GR, being the case of the EH electrodynamics considered in this paper. Moreover, for the quadratic Lagrangian (\ref{f(R)}) the above equation yields $R=-\kappa^2 T$, which is the same relation as in GR.

The gravitational field equations (\ref{eq:eomRBG}) in $ f(R) $ gravity boil down to
\begin{equation}\label{Eq:Ricci_tensor-f(R)}
{R^\mu}_\nu (q) = \dfrac{1}{f_R^2}\left( \dfrac{f}{2} {\delta^\mu}_\nu+ \kappa^2 {T^\mu}_\nu \right) \ ,
\end{equation}
while the deformation matrix in this case becomes ${\Omega^\mu}_{\nu}=f_R \, \delta^\mu_\nu$. Therefore, from (\ref{eq:Omegadef}) the space-time metric $ g_{\mu\nu} $ is conformally related to the Einstein frame metric $ q_{\mu\nu} $ as
\begin{equation}\label{Eq:metrics_fr}
q_{\mu\nu} = f_R \, g_{\mu\nu} \ ,
\end{equation}
where we recall that $f_R \equiv f_R(T)$. Let us proceed with the resolution of the field equations (\ref{Eq:Ricci_tensor-f(R)}). To work as general as possible, at this stage we shall not impose constraints upon the shape of the function $\varphi(X)$. We begin by considering a static, spherically symmetric line element for the $ q_{\mu\nu} $ geometry as
\begin{equation}\label{Eq:line_element_q}
ds^2_q = -A(x) e^{2\psi(x)} dt^2 + \dfrac{dx^2}{A(x)}  +x^2 d\Omega^2,
\end{equation}
where $\psi(x), \, A(x)$ are the two metric functions, and $d\Omega^2=d\theta^2 + \sin^2 \theta d\phi^2$ is the volume element in the unit two-spheres. Now, using the symmetry in $2\times 2$ blocks of the stress-energy tensor (\ref{stress-energy tensor}), the combination $ {R^t}_{t}- {R^x}_{x}=0 $ of the field equations (\ref{Eq:Ricci_tensor-f(R)}) allows to set $\psi(x)=0$ in (\ref{Eq:line_element_q}) without any loss of generality. Now, defining the usual mass ansatz
\begin{equation}\label{Eq:ansatz}
A(x) = 1- \dfrac{2M(x)}{x} \ ,
\end{equation}
we plug it into the remaining non-vanishing component of the field equations as
\begin{equation}
{R^\theta}_{\theta}=\frac{1}{x^2}\left(1-A-xA_x\right)=\frac{2M_x}{x^2} \ .
\end{equation}
Equaling it to the right-hand side of the field equations (\ref{Eq:Ricci_tensor-f(R)}) we find that the mass function satisfies
\begin{equation} \label{eq:Mx}
M_x=\frac{x^2}{2f_R^2} \left(\frac{f}{2}+\frac{\kappa^2 \varphi}{8\pi}\right) \ .
\end{equation}
We next need to express this function in terms of the radial coordinate of the space-time metric $ g_{\mu \nu} $, the latter having the line element
\begin{equation} \label{eq:lineg}
ds^2=-C(x)dt^2+\frac{dx^2}{B(x)} +r^2(x)d\Omega^2 \ ,
\end{equation}
with $C$ and $B$ new functions to be determined.
Eq.(\ref{Eq:metrics_fr}) tells  us that the relation between the radial coordinates on both frames is given by
\begin{equation} \label{eq:xrfR}
x^2=r^2f_R \ .
\end{equation}
Taking a derivative here with respect to $r$ and inserting the result in (\ref{eq:Mx}) we arrive at
\begin{equation}\label{Eq:diff_mass_fr}
M_r = \dfrac{r^2}{4 \,f_R^{3/2}} \left( f + \dfrac{\kappa^2 \varphi}{4 \pi}\right) \left(f_R +\dfrac{r}{2} f_{R,\,r}\right) \ .
\end{equation}
Moreover, by using again (\ref{Eq:metrics_fr}) in the temporal and radial sectors, we arrive to the solution of the line element (\ref{eq:lineg}) as
\begin{equation} \label{eq:finalsolf(R)}
ds^2=-C(x) dt^2 + \frac{dx^2}{f_R^2 C(x)} +z^2(x)d\Omega^2 \ ,
\end{equation}
where we have introduced the dimensionless radial function $z(x)$, which is implicitly defined via Eq.(\ref{eq:xrfR})\footnote{In an abuse of notation, here we have introduced an implicit factor $r_c$ inside $x$ as $x \to x \,r_c$, so Eq.(\ref{eq:xrfR}) reads $x^2=z^2 f_R$.}, while the function $C(x)$ can also be conveniently written in terms of this radial coordinate function as
\begin{equation}\label{eq:Final-ansatz-f(R)}
C(z) = \frac{1}{f_R} \left(1- \dfrac{1 + \delta_1 G(z)}{\delta_2 \, z \, f_R^{1/2}}\right).
\end{equation}
In this metric function we have introduced the two main constants characterizing this problem as
\begin{eqnarray}
\delta_1&=&\frac{r_c^3}{2\,l_{\beta}^2 \,r_S}=\frac{(54\pi)^{3/4}}{2\,r_S} \sqrt{\frac{r_q^3}{l_{\beta}}}  \\
\delta_2&=&\frac{r_c}{r_S} \ ,
\end{eqnarray}
where $r_S \equiv 2M_0 $ is Schwarzschild's radius, while the function $G(z)$ in (\ref{eq:Final-ansatz-f(R)}) is obtained in terms of its derivative as
\begin{equation}\label{eq:G(z)}
G_z (z)=  \frac{z^2}{f_R^{3/2}}\left(\tilde{f} + \dfrac{\kappa^2 \tilde{\varphi}}{4 \pi} \right)\left(f_R + z\,f_{R, \,z}\right) \ ,
\end{equation}
and has contributions from the $ f(R) $ sector as
\begin{eqnarray}
 \tilde{f} &\equiv& l_\beta^2 \,  f= \dfrac{2}{9 \pi} \tau^4 (z) \left(1+\dfrac{\tilde{\alpha}}{2}\tau^4(z)\right)\\
 f_R &=& 1 + \tilde{\alpha}\,  \tau^4(z) \ , \label{eq:fRsol} \label{eq:fRex}
\end{eqnarray}
where $ \tilde{\alpha} \equiv 4\alpha/(9 \pi l_\beta^2) $, and from the NED model as
\begin{eqnarray}
	\tilde{\varphi} &\equiv & l_\beta^2 \, \varphi=\frac{\tau^2(z)}{6\pi}\left(1+\frac{2}{3}\tau(z)\right) \\
	\tau(z) &=& \text{Sinh} \ h(z) \label{eq:tau}\\
	h(z) &=& \frac{1}{3} \ln \left[\frac{1}{z^2}\left(1+ \sqrt{z^4+1}\right)\right] \label{eq:hz} \ .  \label{funcion-zc}
\end{eqnarray}
In the last set of equations, we have introduced the EH model in the characterization of the function $\tau(z)$ (which is just the square root of $X(z)$ in Eq.(\ref{eq:Xz}) removing the constants). The line element (\ref{eq:finalsolf(R)}) with the definitions above is the solution to the problem of electrostatic solutions in quadratic Palatini $f(R)$ gravity coupled to EH electrodynamics, characterized by two integration constants: the mass $M$ and the electric charge $q$; and two new scales:  the gravity parameter $\alpha$ and the matter parameter $\beta$, all of such constants encoded in the two parameters $\delta_1,\delta_2$. We also point out that this line element can be alternatively cast in a more Schwarzschild-like fashion by introducing the change of coordinates $ d\tilde{x}^2= f_R^{-2} dx^2$, though we shall not take this path in order not to spoil the simple representation of the function $z(x)$ in Eq.(\ref{eq:xrfR}).

In the asymptotic limit, $z \to \infty$, one can verify that $f_R \approx 1+\frac{\tilde{\alpha}}{81z^8}$ while the function $G_z$ in Eq.(\ref{eq:G(z)}) boils down to
\begin{equation}
G_z \approx \frac{1}{54\pi z^2} - \frac{1}{729\pi z^6} + \mathcal{O}\left(\frac{\alpha}{z^{10}} \right) \ ,
\end{equation}
which inserted in the metric function (\ref{eq:Final-ansatz-f(R)}) and after spelling out the constants $\delta_1,\delta_2$ yields the result
\begin{equation} \label{eqAsymp}
C(r)  \underset{r \to \infty}{\approx} 1-\frac{r_S}{r}+\frac{q^2}{r^2}-\frac{\beta q^4}{5r^6} + \mathcal{O}\left(\frac{\alpha}{z^{10}}\right) \ ,
\end{equation}
where we have taken units such that $\kappa^2=8\pi$.
The first three terms in this expression correspond to the Reissner-Nordstr\"om black hole of GR, while the next term is the correction from EH electrodynamics. The corrections introduced by $f(R)$ gravity appear only at order tenth, which should not come as a surprise, since the new gravitational dynamics encoded in the theory arise only in the innermost region of the solution, as we shall see next.

\subsection{Properties of the solution: radial function}

\begin{figure}[t]
	\centering
	\includegraphics[width=0.45\textwidth]{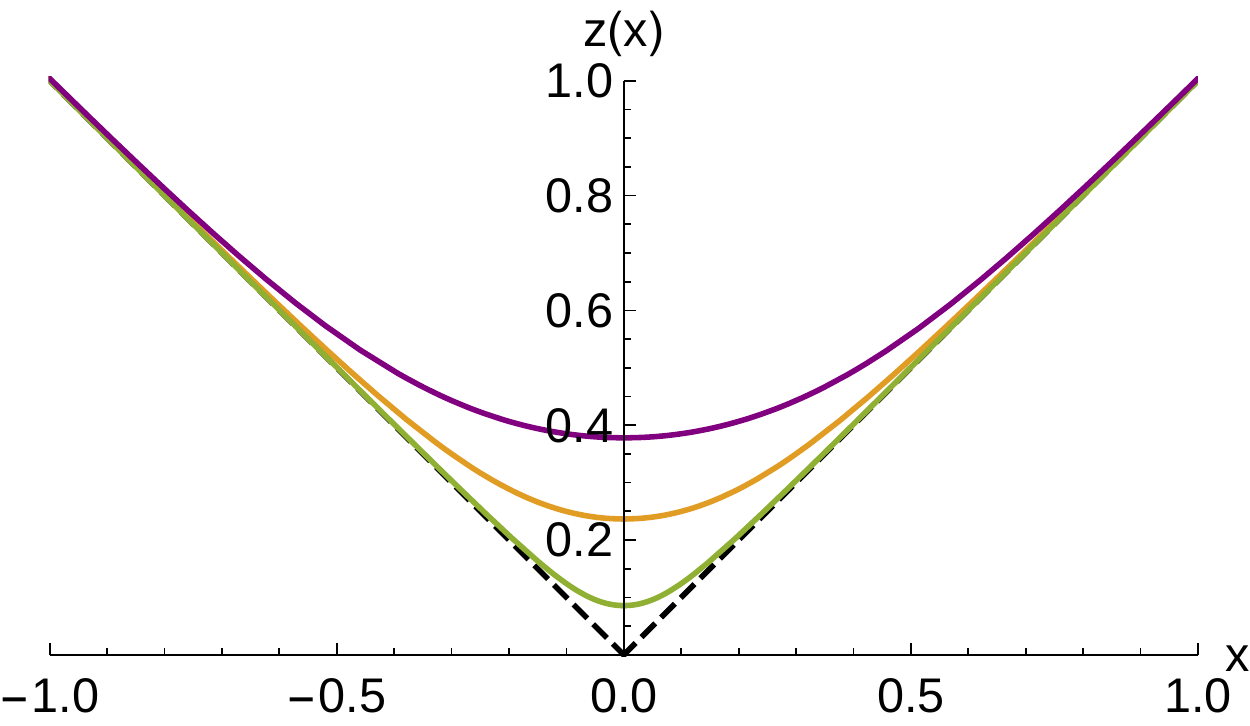}
	\caption{The radial function $z(x)$ in Eq.(\ref{eq:xrfR}) for the case $\tilde{\alpha}<0$ and values $\vert \tilde{\alpha} \vert = 0.2 $ (green), $\vert \tilde{\alpha} \vert = 0.5 $ (orange) and $\vert \tilde{\alpha} \vert = 1.0 $ (purple). The wormhole throat is located at $x=0$ ($z=z_c$ as defined in Eq.(\ref{eq:zcfR})). As a comparison, we have plotted the GR case, $r^2=x^2$ (black dashed), for which no such a bounce in the radial function is present.}
	\label{fig:rfrneg}
\end{figure}

We begin now our analysis of the most relevant features of these solutions by considering the behaviour of the radial function $z\, (x)$ in (\ref{eq:xrfR}). From (\ref{eq:fRex}) and the positivity everywhere of $\tau\,(z)$, one finds that for $\tilde{\alpha}<0$ the function $f_R$ will vanish at a certain $z=z_c$ with
\begin{equation} \label{eq:zcfR}
z_c=\sqrt{\frac{2\,a}{a^2-1}} \ ,
\end{equation}
where we have introduced the new constant
\begin{equation}
a=\exp \left\lbrace 3\,\text{ArcSinh}\left(\vert \tilde{\alpha} \vert^{-1/4}\right)\right\rbrace \ .
\end{equation}
Unfortunately, Eq.(\ref{eq:xrfR}) does not admit a closed expression for $z=z\,(x)$ in its full domain of definition, but it is easy to see that at $z=z_c$ one has $x=0$ and beyond this point the radial function bounces off to another asymptotically flat region of space-time (see Fig.\ref{fig:rfrneg}). Therefore, the area of the two-spheres $S=4\pi z^2$ is bounded from below, and the space-time consist of two patches of the radial function $z \in [z_c,\infty)$ or a single one in terms of the radial coordinate $x \in (-\infty,+\infty)$.
The natural interpretation for this bouncing behavior and minimum areal function is that of a wormhole structure \cite{VisserBook}, with $z=z_c$ representing its throat. The size of the latter grows with $\vert \tilde{\alpha} \vert$, while it closes in the limit $\vert \tilde{\alpha} \vert \to 0$, corresponding to GR.

\begin{figure}[t]
	\centering
	\includegraphics[width=0.45\textwidth]{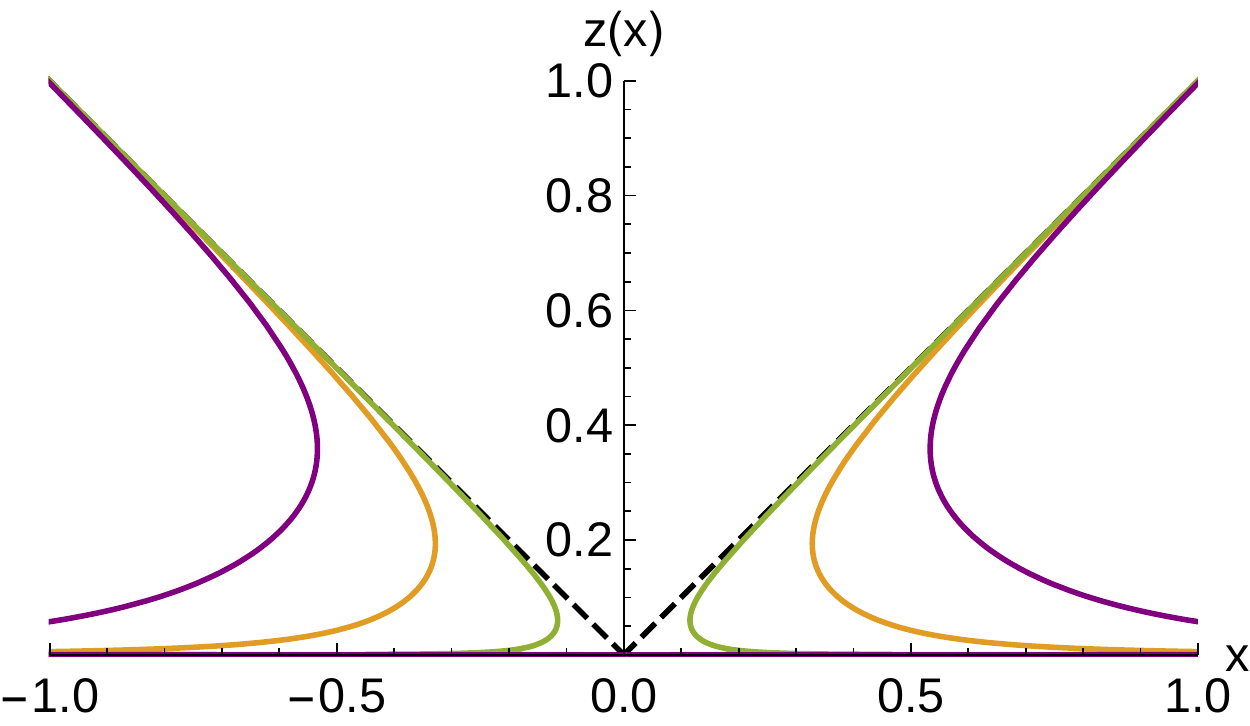}
	\caption{Plot of the dimensionless coordinate $ z $ as a function of $ x $ for the case $ \tilde{\alpha} >0 $. Same notation as in Fig. \ref{fig:rfrneg}. }
	\label{fig:rfrpos}
\end{figure}

In the $ \tilde{\alpha}>0 $ case, things are far less interesting. Indeed, in such a case $f_R$ has not zeros and the radial function $z(x)$ does not yield a bounce, but instead generates two branches of solutions. As depicted in Fig.\ref{fig:rfrpos}, there is not a smooth transition between these two branches, and the area of the two-spheres can go all the way down to vanishing value. The corresponding solutions are presumably singular and, therefore, we shall no longer consider them here.

\subsection{Properties of the solution: inner behaviour and horizons}

\begin{figure}[t!]
	\centering
	\includegraphics[width=0.41\textwidth]{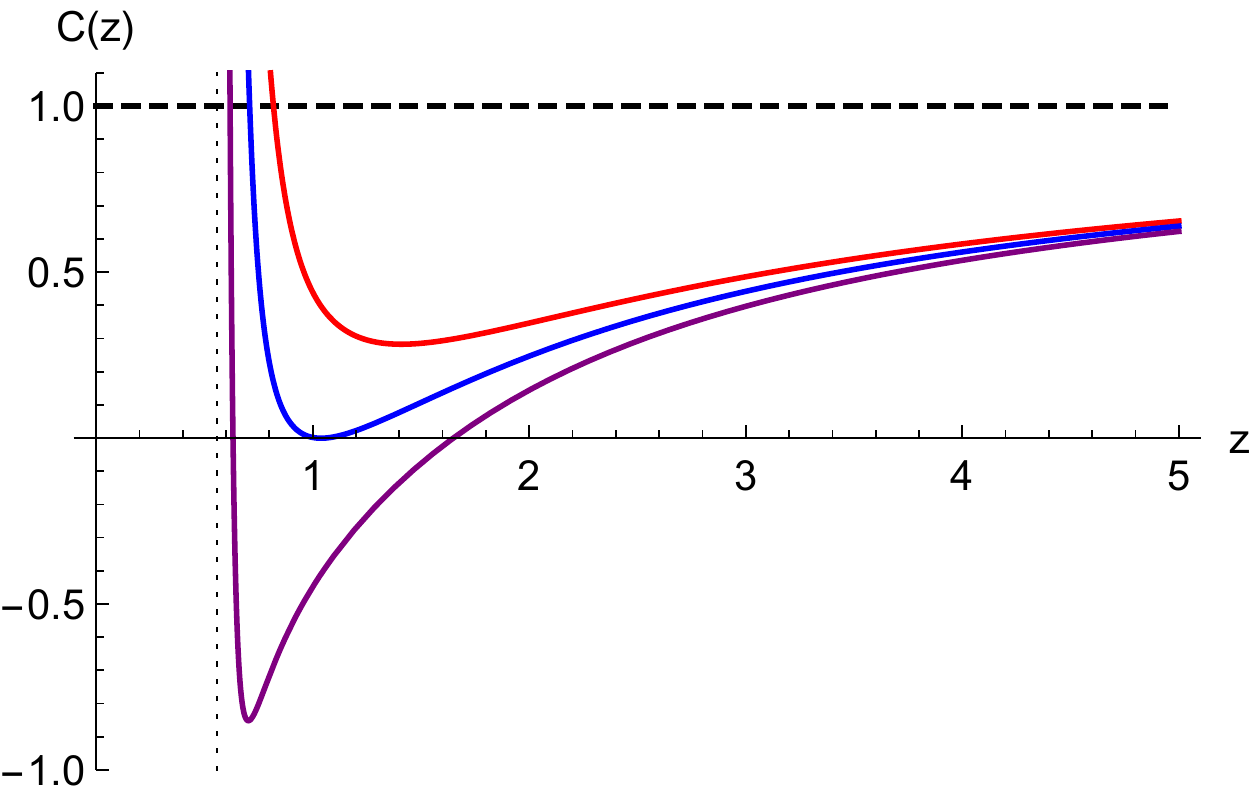}
	\caption{Numerical integration of the metric function $C(z)$ in Eq.(\ref{eq:Final-ansatz-f(R)}) for the branch $\tilde{\alpha}<0$ and the choices of $\vert \tilde{\alpha} \vert=5, \delta_2=1/2$ and three values of $\delta_1=50$ (purple), $\delta_1 \approx 85$ (blue) and  $\delta_1=120$ (red), corresponding to black holes with two horizons, extreme black holes, and naked configurations, respectively. All solutions are asymptotically flat, $C(z) \underset{z \to \infty}{\approx} 1$. The vertical dotted line represents the wormhole throat $z=z_c$, to which all curves converge.}
	\label{fig:met1}
\end{figure}

From now on we shall focus on characterizing the properties of the branch $\tilde{\alpha}<0$\footnote{Since $\beta>0$ in order for the weak energy condition on EH electrodynamics to be satisfied, then this branch implies that the gravity coupling constant in the action (\ref{f(R)}) must satisfy $\alpha<0$. Note in this sense that no physical constraint forbids the curvature corrections to become positive/negative in the ultraviolet limit.}, where we have some hope of finding regular black hole solutions. Since the deviations as compared to GR solutions are expected to arise in the innermost region, we need to study the behaviour of the functions $f_R$ and $G(z)$ there. In this sense, a series expansion of the former using (\ref{eq:fRsol}) around $z=z_c$, as defined in (\ref{eq:zcfR}), yields the result
\begin{equation}\label{Eq:fR_expansion}
f_R \approx f_1
 (z-z_c) + \mathcal{O}(z-z_c)^2 \ ,
\end{equation}
where we have introduced the constant
\begin{equation}
f_1(z_c)= \frac{8  \,\text{coth} \ h(z_c)}{3 \,z_c \, \sqrt{z_c^4+1}} \ ,
\end{equation}
with $h(z)$ defined in Eq.(\ref{eq:hz}). As for $G_z$, it turns out to be a tougher nut to crack given its involved functional dependence, but it behaves at $z=z_c$ as
\begin{equation}
G_z\approx \frac{C_1(z_c)}{(z-z_c)^{3/2}} + \mathcal{O}(z-z_c)^{-1/2} \ ,
\end{equation}
where $C_1(z)>0$ is a cumbersome function of the constants of the model. Therefore, $G(z)\approx \frac{-2C(z_c)}{(z-z_c)^{1/2}} + \mathcal{O}(z-z_c)^{1/2}$, which inserted into the expression for the metric function (\ref{eq:Final-ansatz-f(R)}) yields
\begin{equation} \label{eq:CzcfR}
C(z) \approx \frac{\tilde{C}_1(z_c) \delta_1}{\delta_2 (z-z_c)^2} + \mathcal{O}(z-z_c)^{-3/2} \ ,
\end{equation}
where the new constant $\tilde{C}_1(z)>0$ contains all the contributions in $z_c$. Therefore, we see that the metric function $C(z)$ diverges always at $z=z_c$, as a consequence of the poles present in the $f_R$ factor and also in the $G(z)$ function. Moreover, due to the positivity of $\tilde{C}_1(z), \delta_1$ and $\delta_2$ in this expression, one finds that this divergence goes always to $+\infty$ which, together with the asymptotically flat character of the solutions, as given by (\ref{eqAsymp}), provides the structure of horizons for these solutions. Indeed, as depicted in Fig.\ref{fig:met1}, this structure resembles the one of the Reissner-Nordstr\"om solution of GR, namely, two-horizon black holes, extreme black holes (with a single degenerated horizon) and naked configurations. However, a systematic classification of the values of $\{\vert \tilde{\alpha} \vert, \delta_1, \delta_2\}$ yielding any such configurations are hard to find, and require instead direct inspection case-by-case. We also see that the single horizon black holes of the EH electrodynamics in GR have been lost, due to the modifications on the geometry caused by the presence of the wormhole throat at a finite distance $z_c$.

\subsection{Properties of the solution: geodesic behaviour and regularity}

\begin{figure}[t!]
	\centering
	\includegraphics[width=0.45\textwidth]{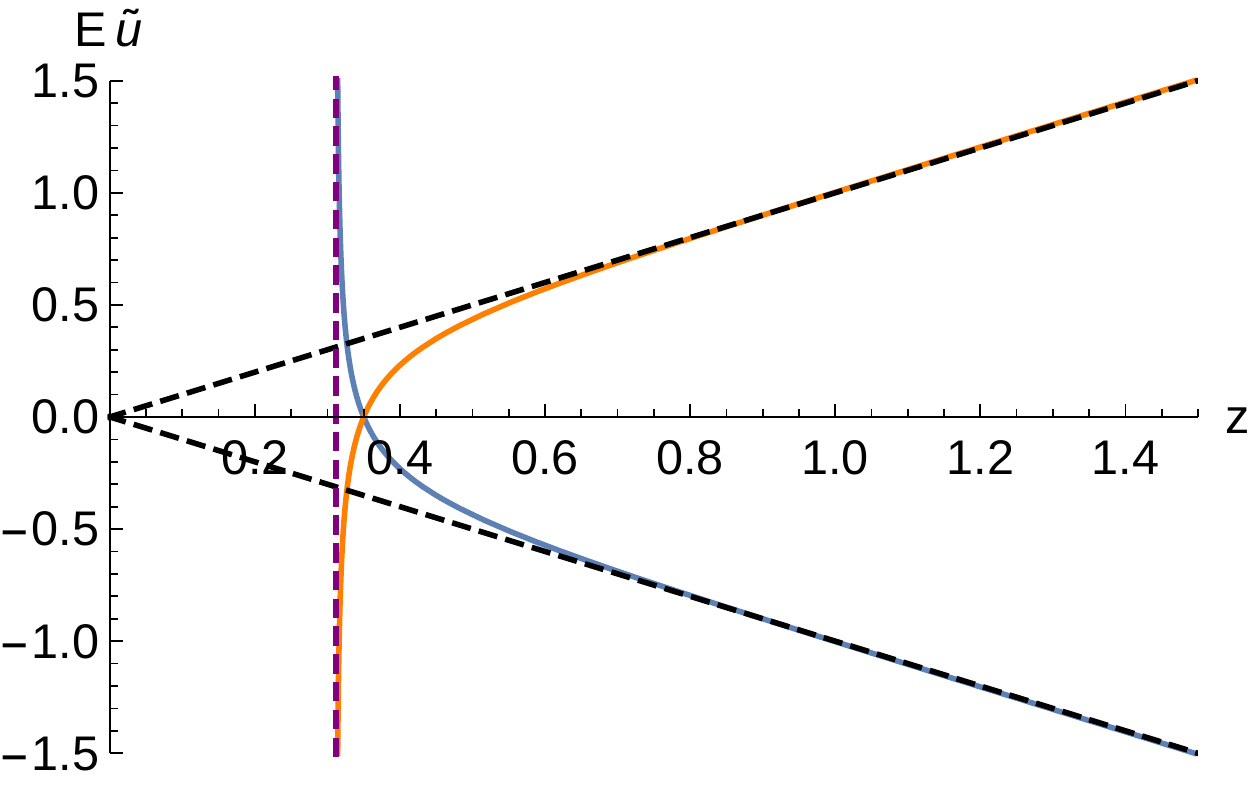}
	\caption{The affine parameter $ E \cdot \tilde{u}(z) $ versus the dimensionless radial coordinate $ z $ for ingoing (blue) and outgoing (orange) null radial geodesics. The vertical dashed purple line corresponds to the wormhole throat, $z=z_c$, while the black dashed lines correspond to null radial geodesics in GR.}
	\label{fig:geofr}
\end{figure}

To gain deeper knowledge on the innermost geometry of these solutions let us study their geodesic structure. For any spherically symmetric space-time with line element (\ref{eq:lineg}) the geodesic equation may be written as \cite{OlmoBook}
\begin{equation}\label{Eq:General_geo}
	\dfrac{C}{B}\left( \dfrac{dx}{du}\right)^2 = E^2 -V(x) \ ,
\end{equation}
where we have introduced the effective potential
\begin{equation}\label{Eq:Potential_f(R)}
V(x) = C(x) \left( -k + \dfrac{L^2}{r^2(x)}\right) \ .
\end{equation}
Here, $u$ is the affine parameter (the proper time for a time-like observer), $k=-1,0$ for time-like and null geodesics, respectively, while $ E $ and $L  $ are the total energy and angular momentum per unit of mass for time-like observers, respectively.  For spherically symmetric space-times in Palatini $ f(R) $ gravity, from (\ref{eq:finalsolf(R)}) the geodesic equation (\ref{Eq:General_geo}) takes the form
\begin{equation}\label{Eq:General_geo_f(R)}
	\dfrac{1}{f_R^2 }\left( \dfrac{dx}{du}\right)^2 = E^2 -V(x) \ .
\end{equation}
It is convenient to rewrite the above equation in terms of the dimensionless radial function $z(x)$ by using Eq. \eqref{eq:xrfR} and its derivatives. This allows to write the following form of the geodesic equation
\begin{equation}\label{Eq:Part_geo_f(R)}
\dfrac{d\tilde{u}}{dz} = \pm \dfrac{ 1+\frac{z \, f_R,_z}{2 \, f_R} }{f_R^{1/2}\sqrt{E^2 -C(z)  \left( -k + \frac{L^2}{r_c^2 z^2(x)}\right)}} \ ,
\end{equation}
where we have re-scaled the affine time parameter as $ \tilde{u} = u/r_c $, and the +(-) sign come from taking the square-root in (\ref{Eq:General_geo_f(R)}) and correspond to ingoing (outgoing) geodesics, that is, trajectories of particles leaving (getting to) the wormhole throat $z=z_c$. For radial null geodesics ($ k=0 $ and $ L=0 $), the above differential equation becomes
\begin{equation}\label{Eq:null_f(R)}
E \,\dfrac{d\tilde{u}}{dz} = \pm \dfrac{ 1+\frac{z \, f_R,_z}{2 \, f_R} }{f_R^{1/2} } \ .
\end{equation}
At large distances, $ z \gg 1 $, where $f_R \to 1$, this equation can be integrated as $ E \tilde{u}\simeq \pm z $, which is the expected GR behaviour, in agreement with the fact that in this limit the $f(R)$-EH solution boils down to the Reissner-Nordstr\"om one.  However, departures are expected as the wormhole throat $z=z_c$ is approached. Indeed, using the expansion of $ f_R $ in Eq. \eqref{Eq:fR_expansion} we can easily integrate this expression around $z=z_c$ as
\begin{equation}\label{Eq:critical_geo_f(R)}
 E \tilde{\lambda}(z)\approx \mp \frac{\sqrt{8/3}\,z_c}{f_1(z_c)} \frac{1}{\sqrt{z-z_c}} \ .
\end{equation}
From this expression it is readily seen that the affine parameter $\tilde{\lambda}(z)$ diverges to $\pm \infty$ as the wormhole throat $z=z_c$ is approached (see Fig. \ref{fig:geofr}). This implies that null radial geodesics require an infinite affine time  to get to (or to depart from) the wormhole throat which, consequently, lies on the future (or past) boundary of the space-time\footnote{In some sense this means that we have {\it half a wormhole}, in that the region $x>0$ ($x<0$) is not accessible to observers living in the region $x<0$ ($x>0$).}. This way, as opposed to the Reissner-Nordstr\"om space-time where null radial geodesics get to $r=0$ in finite affine time without any possibility to further extension beyond this point, they are complete in the geometry explored in this section. We point out that this mechanism for the removal of geodesic incompleteness via the displacement of any potentially pathologically region to the boundary of the space-time has been discussed in detail in Refs.\cite{Carballo-Rubio:2019fnb,Carballo-Rubio:2019nel} on very general grounds, and explicitly implemented in other settings within Palatini theories of gravity \cite{Bambi:2015zch,Bejarano:2017fgz,Nascimento:2018sir}.

For null geodesics with $L \neq 0$ and for time-like geodesics, the fact that $C(x)$ diverges to $+\infty$ at $z=z_c$, implies that any such geodesics approaching the wormhole throat will see an infinitely repulsive potential barrier, as given by (\ref{Eq:Potential_f(R)}), and will bounce off at a certain radius given by the vanishing of the denominator of (\ref{Eq:Part_geo_f(R)}), thus not being able to get to the wormhole throat. Consequently, these geodesics are complete in pretty much the same way as in their Reissner-Nordst\"om counterparts. The bottom line of this discussion is the null and time-like geodesic completeness of the full spectrum of solutions (in terms of mass and charge) of quadratic $f(R)$ gravity with EH electrodynamics in the $\tilde{\alpha}<0$ branch. Regarding the behaviour of curvature at the wormhole throat, a quick computation revels the existence of curvature divergences there of size $K \equiv {K^\alpha}_{\beta\gamma\delta}{K_\alpha}^{\beta\gamma\delta} \sim 1/(z-z_c)^2$, which are nonetheless much weaker than in their GR counterparts $K \sim 1/r^8$. One could wonder what is the meaning of such divergences, since neither geodesics can reach such a region nor information can come out of it to affect asymptotic observers\footnote{There is the issue, though, about whether the presence of unbound curvature in such a region might spoil the  well-behaved Cauchy development of past null infinity, which would require further research beyond the one carried out in this work.}.

\section{EiBI gravity}  \label{sec:IV}

\subsection{Derivation of the solution}

The action of EiBI gravity can be written as (for a review of this theory see \cite{BeltranJimenez:2017doy})
\begin{equation}
\mathcal{S}_{EiBI}=\frac{1}{\kappa^2 \epsilon} \int d^4x \left[\sqrt{-\vert g_{\mu\nu} + \epsilon R_{\mu\nu} \vert} - \lambda \sqrt{-g}\right] \ ,
\end{equation}
where $\epsilon$ is a parameter with dimensions of length squared\footnote{The size of this parameter in both its positive/negative branches has been the subject of many studies in the literature, see \cite{Olmo:2019flu} for details on astrophysical constraints.}.
In its weak-field limit, $\vert R_{\mu\nu} \vert \ll \epsilon^{-1}$, the theory reduces to GR with an effective cosmological constant, $\Lambda_{eff}=\frac{\lambda-1}{\epsilon}$, while higher order curvature corrections are suppressed by powers of $\epsilon$. For EiBI gravity, the Einstein-frame metric appearing in (\ref{eq:Omegadef}) is given by $q_{\mu\nu} =g_{\mu\nu} + \epsilon R_{\mu\nu}$, while the deformation matrix  $ {\Omega^\mu}_{\nu} $ can be determined via the algebraic expression
\begin{equation}\label{Eq:def-Omega}
|\hat{\Omega}|^{1/2} ({\Omega^\mu}_{\nu})^{-1} = \lambda \, {\delta^\mu}_{\nu}- \epsilon \kappa^2 {T^\mu}_{\nu} \ .
\end{equation}
This relation shows that the deformation matrix inherits also in this case the structure in  $2 \times 2$ blocks of the stress-energy tensor defined in Eq. \eqref{stress-energy tensor}. Thus, we are allowed to write an ansatz for $ {\Omega^\mu}_{\nu} $ as
\begin{equation}\label{Eq:Omega}
{\Omega^\mu}_{\nu} = \left(\begin{array}{cc} \Omega_+ \; \hat{I}_{2\times 2}&  \hat{0\,}_{2\times 2}\\  \hat{0\,}_{2\times 2}&  \Omega_- \;  \hat{I}_{2\times 2}\end{array}\right) \ ,
\end{equation}
where the components of the matrix can be found by substituting them into Eq.\eqref{Eq:def-Omega} and solving the corresponding equations as
\begin{eqnarray}
\Omega_+ &=& \lambda-\epsilon \kappa^2 {T^\theta}_{\theta}= \lambda - \dfrac{\epsilon \kappa^2}{8 \pi} \varphi \ , \label{Omplusgen} \\
\Omega_- &=& \lambda-\epsilon \kappa^2 {T^t}_{t}= \lambda - \dfrac{\epsilon \kappa^2}{8 \pi} (\varphi- 2X \varphi_X)  \label{Omneggen} \ .
\end{eqnarray}
The gravitational field equations in this case are written as
\begin{equation}\label{Eq:Ricci_tensor-EIBI}
{R^\mu}_\nu (q) = \dfrac{1}{\epsilon} \left( \begin{array}{cc} \dfrac{\Omega_+-1}{\Omega_+} \; \hat{I}_{2\times 2}&  \hat{0}_{2\times 2}\\  \hat{0}_{2\times 2}&  \dfrac{\Omega_- -1 }{\Omega_-}\;  \hat{I}_{2\times 2}\end{array} \right).
\end{equation}
Considering the line element in \eqref{Eq:line_element_q} and following the same steps done in the previous section, besides taking into account  the relation (\ref{eq:Omegadef}) between metrics, which implies the following relation between radial coordinates
\begin{equation} \label{eq:xrBI}
x^2=z^2 \Omega_-(z) \ ,
\end{equation}
together with Eq.(\ref{Eq:def-Omega})
leads to the expression for the mass function
\begin{equation}\label{Eq:diff_mass_eibi}
M_r= \dfrac{r^2 }{2\, \epsilon }\,(\Omega_--1)\, {\Omega_-}^{1/2}\left( 1 + \dfrac{r \, \Omega_{-,r}}{2 \; \Omega_-^{1/2}} \; \right) \ .
\end{equation}
Moreover, following a similar procedure and notation as in the $ f(R) $ gravity case, we find the line element for the $ g_{\mu \nu} $ metric as
\begin{equation} \label{eq:finalsolEiBI}
ds^2=-C(x) \, dt^2 + \frac{dx^2}{\Omega_{+}^2 \,C(x)} +z^2(x)\,d\Omega^2 \ ,
\end{equation}
where the metric function now satisfies
\begin{equation}\label{eq:Final-ansatz-EiBI}
C(z) = \dfrac{1}{\Omega_+}\left( 1- \dfrac{1 + \delta_1 G(z)}{\delta_2 \, z \, \Omega_-^{1/2}}\right) \ .
\end{equation}
Here, we have introduced the following definitions: the metric is parameterized in terms of two constants  defined as
\begin{eqnarray}
 \delta_1 &=& \dfrac{r_c^3}{r_S \, \epsilon} = \dfrac{3}{2}\left(\dfrac{3}{2 \, \pi}\right) ^{1/4}\frac{1}{  l_\epsilon^2\,r_S} \sqrt{\frac{r_q^3}{l_{\beta}}} \ ,  \\
 \delta_2 &=& \dfrac{r_c}{r_S} \ ,
\end{eqnarray}
where we have redefined the EiBI parameter as $ l_{\epsilon}^2=   \epsilon\, /(12 \pi \, l_\beta^2) $, which is the analog of $ \tilde{\alpha} $ in the $ f(R) $ case. These two parameters encode the two integration constants, $r_S$ and $r_q^2$, and the two gravity and model parameters, $l_{\epsilon}^2$ and $l_{\beta}^2$, likewise in the $f(R)$ case. As for  the $ G(z) $ function, it is obtained via
\begin{eqnarray} \label{eq:GzBI}
G_z(z) &=& z^2 (\Omega_--1)\; \Omega_-^{1/2}\left(1+ \dfrac{z  \Omega_{-},{z}}{2 \, \Omega_-} \;\right). \label{Eq:G(z)_EiBI}
\end{eqnarray}
and the contributions on the $\Omega_{\pm}$ factors read
\begin{eqnarray}
\Omega_{+} &=& \lambda -l_{\epsilon}^2 \,\tau^2(z) \left(1+\frac{ 2\,\tau^2(z)}{3} \right) \ , \label{eq:Op-bi} \\
\Omega_{-} &=& \lambda + l_{\epsilon}^2\,\tau^2(z) \,(1+2\,\tau^2(z)) \ , \label{eq:Om-bi}
\end{eqnarray}
where we recall that $\tau(z)$ is defined in Eq.(\ref{eq:tau}). The line element written in \eqref{eq:finalsolEiBI} is the electrostatic solution of EiBI gravity coupled to EH electrodynamics. Like in the $f(R)$ case, one could transform the line element (\ref{eq:finalsolEiBI}) into a Schwarzschild-like form via the change of coordinates $d\tilde{x}^2=\frac{dx^2}{\Omega_{+}^2}$, but again we shall not follow that path in order not to spoil the simple representation (\ref{eq:xrBI}) of the radial function. From now on, we will consider asymptotically flat solutions, $ \lambda=1 $.

Let us now first analyze the asymptotic limit of the functions in the line element (\ref{eq:finalsolEiBI}). For $ z\rightarrow\infty $, one has $z^2 \approx x^2$, and
the deformation metric components behave as
\begin{equation}\label{eq:def-matr-inf}
\Omega_{\pm} \approx 1\mp \frac{l_{\epsilon}^2}{9 z^4}+\mathcal{O}\left(\frac{1}{z}\right)^8.
\end{equation}
Here, the gravitational sector contributes to the line element in a lower power of $ z $ in comparison to the $ f(R) $ case because $ \Omega_{-} $ has a power in $ \t(z) $. As a consequence, the gravitational corrections appear earlier in the metric component $ C(z) $. Expanding the function $ G_z $ in \eqref{Eq:G(z)_EiBI} we get
\begin{equation}\label{key}
 G_z  \approx\frac{ l_{\epsilon}^2}{9 z^2}-\frac{ \epsilon \, (9  \,l_{\epsilon}^2+4)}{486  z^6}+\mathcal{O}\left(\frac{1}{ z}\right)^{10}.
\end{equation}
Replacing the above expressions into the metric component \eqref{eq:Final-ansatz-EiBI} and reverting back to the original variables leads to
\begin{equation}\label{eqAsymp-EiBI}
C(r)  \underset{r \to \infty}{\approx} 1-\frac{r_S}{r}+\frac{q^2}{r^2}+\dfrac{\epsilon \, r_S q^2}{2 \,r^5}-\frac{(\beta +4\epsilon) \,q^4}{5\,r^6} + \mathcal{O}\left(\frac{1}{r}\right)^{10},
\end{equation}
The first three terms in this expression correspond to the Reissner-Nordstr\"om solution, as expected. The fourth one introduces a sort of interaction between mass and charge fueled by the EiBI gravity dynamics, while the last two terms are pure corrections in EH electrodynamics (obviously identical to the one written in Eq.\eqref{eqAsymp}) and in EiBI gravity, respectively.

\begin{figure}[t!]
	\centering
	\includegraphics[width=0.45\textwidth]{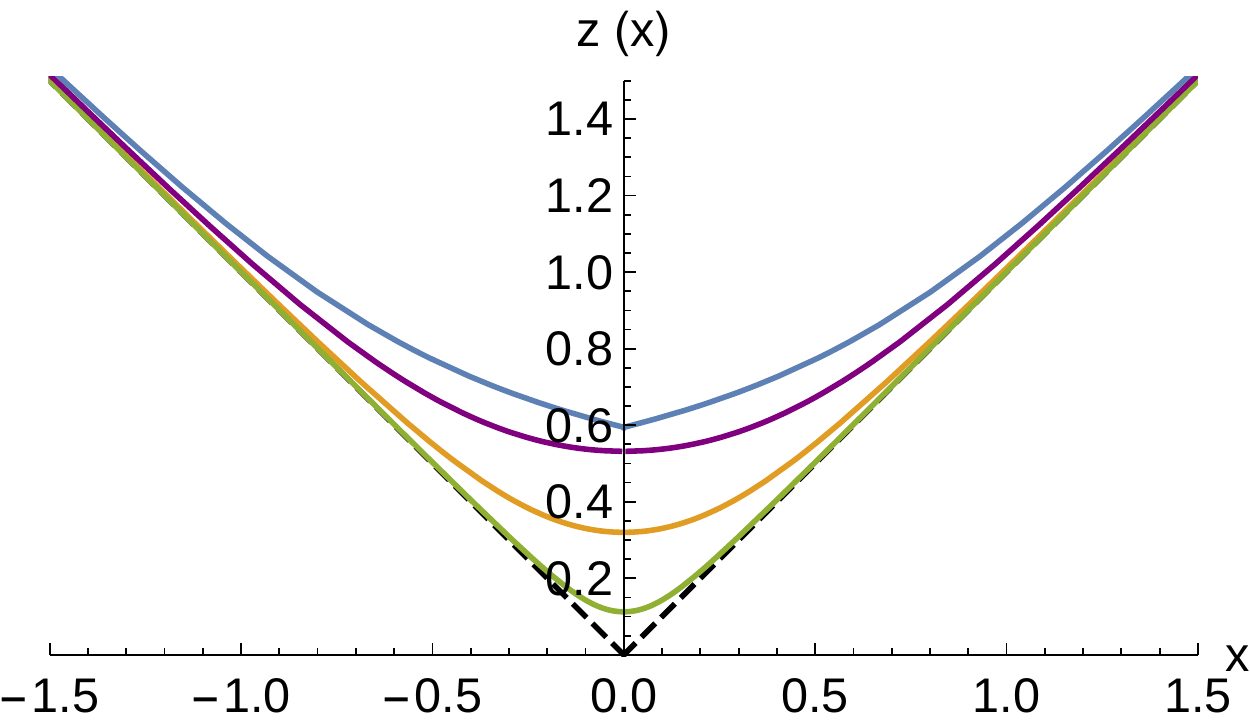}
	\caption{The dimensionless radial function $ z(x) $ for the case $ l_{\epsilon}^2<0$. The orange curve represents $ \vert l_{\epsilon}^2 \vert = 0.2$, the green  $ \vert l_{\epsilon}^2 \vert = 0.01$ and the purple  $ \vert l_{\epsilon}^2 \vert = 1$. The blue and black curves represent the case of Maxwell electrodynamics  with  $ \vert l_{\epsilon}^2 \vert = 0.5$ and $ r_q= 0.5 $, and of GR, respectively.}
	\label{fig:rEiBIneg}
\end{figure}

\subsection{Properties of the solution: radial function}

\begin{figure}[t!]
	\centering
	\includegraphics[width=0.45\textwidth]{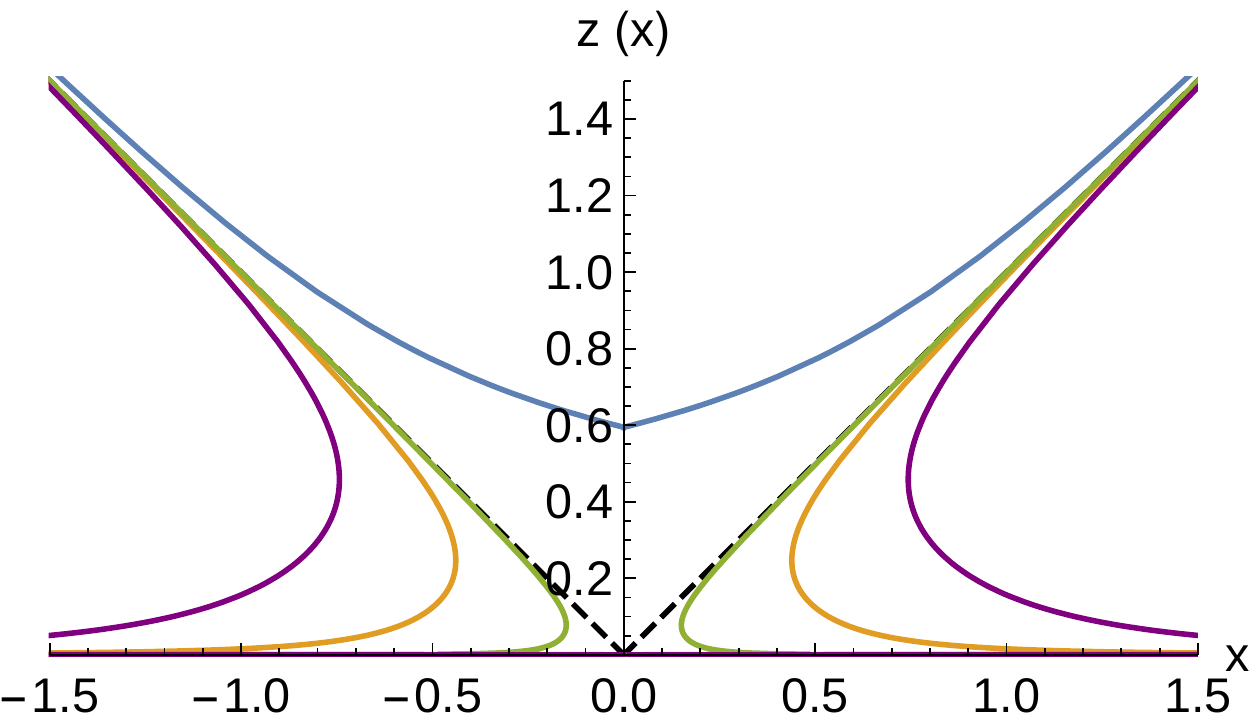}
	\caption{The radial function $ z $ for the case $ l_{\epsilon}^2>0 $. Same notation as in Fig. \ref{fig:rEiBIneg}.}
	\label{fig:rEiBIpos}
\end{figure}

As in the $f(R)$ case, we now look for the minimum of the radial function $z(x)$ via the relation (\ref{eq:xrBI}). Using the expression (\ref{eq:Om-bi}) it is clear that the zeros of $\Omega_{-}$ will only occur in the branch $l_{\epsilon}^2<0$ and, therefore, we shall restrict our attention to this branch from now on. The values of $z(x)$ for which the zeros of $\Omega_{-}$ are attained can actually be written in an identical form as Eq. \eqref{eq:zcfR}, but  now with
\begin{equation}\label{Eq:a-EiBI}
a=\exp\left\lbrace {3\,\text{ArcSinh}\left(\dfrac{1}{2} \sqrt{\sqrt{\dfrac{\vert l_{\epsilon}^2\vert +8}{\vert l_{\epsilon}^2 \vert}}-1}\right)}\right\rbrace.
\end{equation}

As it is depicted in Fig. \ref{fig:rEiBIneg}, at this point the radial function $z(x)$ takes its minimum value and bounces off, again representing a wormhole structure with $z=z_c$ the location of its throat. For completeness, we have also plotted in Fig. \ref{fig:rEiBIpos} the behaviour of $z(x)$ in the branch $ l_\epsilon^2 >0 $, where no wormhole is present and the space-time splits into two disconnected pieces in the $x>0$ and $x<0$ regions. Therefore, these two structures are similar to those found in the $f(R)$ case above, but their effects in the geometry of the corresponding space-times yield large differences, as we shall see next.

\subsection{Properties of the solution: inner behavior and horizons}

To study the behavior of the metric functions on the innermost region and, in particular, at the wormhole throat, we begin by expanding the relevant functions around $z \approx z_c$. For the $\Omega_{\pm}$ functions in (\ref{Omplusgen}) and (\ref{Omneggen}) with the expression (\ref{final-stress-energy tensor}) we find
\begin{eqnarray}
\Omega_{+} &\approx& \omega_{+}(z_c) +\mathcal{O}(z-z_c)\label{Omega_+_zc} \ , \\
\Omega_{-}&\approx&\ \omega_{-}(z_c) (z-z_c)+	\mathcal{O}(z-z_c)^2 \ , \label{Omega_-_zc}
\end{eqnarray}
where we have introduced the constants
\begin{eqnarray}
\omega_{+}(z_c)&=&\frac{2}{3} \left(\sech 2\,h(z_c)+2\right) \ , \label{omega_+_zc} \\
\omega_{-}(z_c)&=&\dfrac{4}{3}\,\frac{(\tanh 2 \,h(z_c)+\coth h(z_c))}{ z_c \sqrt{z_c^4+1}} \ , \label{omega_-_zc}
\end{eqnarray}
and we recall that $ h(z_c) $ is defined in Eq.\eqref{funcion-zc}. The expansion of the function $G_z$ in Eq.(\ref{eq:GzBI}) becomes
\begin{equation}\label{Gz_zc}
G_z \approx \dfrac{ C_2}{\sqrt{z-z_c}}\, +\mathcal{O}(z-z_c)^{1/2} \ ,
\end{equation}
where  the constant $
C_2=z_c^3 \omega_-^{1/2}/2$.
Upon integration, this yields the result
\begin{equation}\label{G(z)_zc}
G(z) \approx -\frac{1}{\delta_c} + 2\, C_2\, \sqrt{z-z_c}+\mathcal{O}(z-z_c)^{3/2} \ ,
\end{equation}
where $\delta_c(z_c)>0$ is a constant needed to match the inner and asymptotic expansions of $G(z)$, and whose explicit dependence on its argument is very cumbersome, though for our analysis only its positivity is relevant. Plugging the expansions (\ref{Omega_+_zc}), (\ref{Omega_-_zc}) and (\ref{G(z)_zc}) in the expression (\ref{eq:Final-ansatz-EiBI}), we arrive at the behaviour of the metric components:
\begin{eqnarray}\label{C-EiBI-zc}
g_{tt} &\approx& - \frac{3 \left(1+2 \tau^2_c\right) (\delta_1/\delta_c-1)}{2\, z_c \delta_2  \left(3+4 \,\tau^2_c\right) \omega_{-}^{1/2} \sqrt{z-z_c}}\\
&+&\frac{3 \left(\delta_2-\delta_1 z_c^2\right)}{2\, \delta_2 (2+\sinh 2\,h(z_c))}+\mathcal{O}(z-z_c)^{1/2} \ , \nonumber \\
g_{rr}&\approx& \frac{3 \,z_c\,\delta_2 \,\omega_{-}^{1/2}\cosh 2\,h(z_c)}{ 2(\delta_1/\delta_c-1)(3+2 \tau_c)} \sqrt{z-z_c}+\mathcal{O}(z-z_c) \ , \label{grr-EiBI-zc}
\end{eqnarray}
where $ \tau_c \equiv \tau (z_c)$.

From these expressions we can proceed to classify the spectrum of solutions in terms of their horizon structure. Indeed, a glance at the expansion (\ref{C-EiBI-zc}) shows that such a classification can be performed according to the ratio $\delta_1/\delta_c$, since it controls the sign of the divergence of the metric function $C(z)$ at $z=z_c$. Thus, if $\delta_1/\delta_c<1$ then $C(z_c) \to -\infty$, and the corresponding solutions are Schwarzschild-like black holes with a single horizon. On the contrary, when $\delta_1/\delta_c>1$, then $C(z_c) \to +\infty$ and one finds configurations with the same structure as the one  of the Reissner-Nordstr\"om solution of GR: black holes with two horizons, extreme black holes, or naked configurations, depending on the value of the constant $\delta_2 $. Moreover, special configurations are found when $\delta_1=\delta_c$ since in such a case, replacing first this constraint in the metric functions of the line element (\ref{eq:finalsolEiBI}) and expanding in series of $z_c$ makes the first term in Eq.(\ref{C-EiBI-zc}) to go away and only the finite contribution at $z=z_c$ remains. Consequently, the corresponding configurations are Minkowski-like solutions with either a single non-degenerate horizon or none, depending on the value of $\delta_2\gtrless \delta_1 z_c^2$. It should be pointed out that the Schwarzschild/Reissner-Nordstr\"om-like structure of horizons resemble the original one of the EH electrodynamics within GR: while in the latter it is the comparison between the total mass of the space-time, $M$, and the total (finite) energy stored in the electrostatic field the one playing the role in classifying such a structure, here is the ratio $\delta_1/\delta_c$ instead. However, the Minkowski-like configurations are a novel feature of these Palatini space-times, having no counterpart in the Einstein-EH system.

\begin{figure}[t!]
	\centering
	\includegraphics[width=0.45\textwidth]{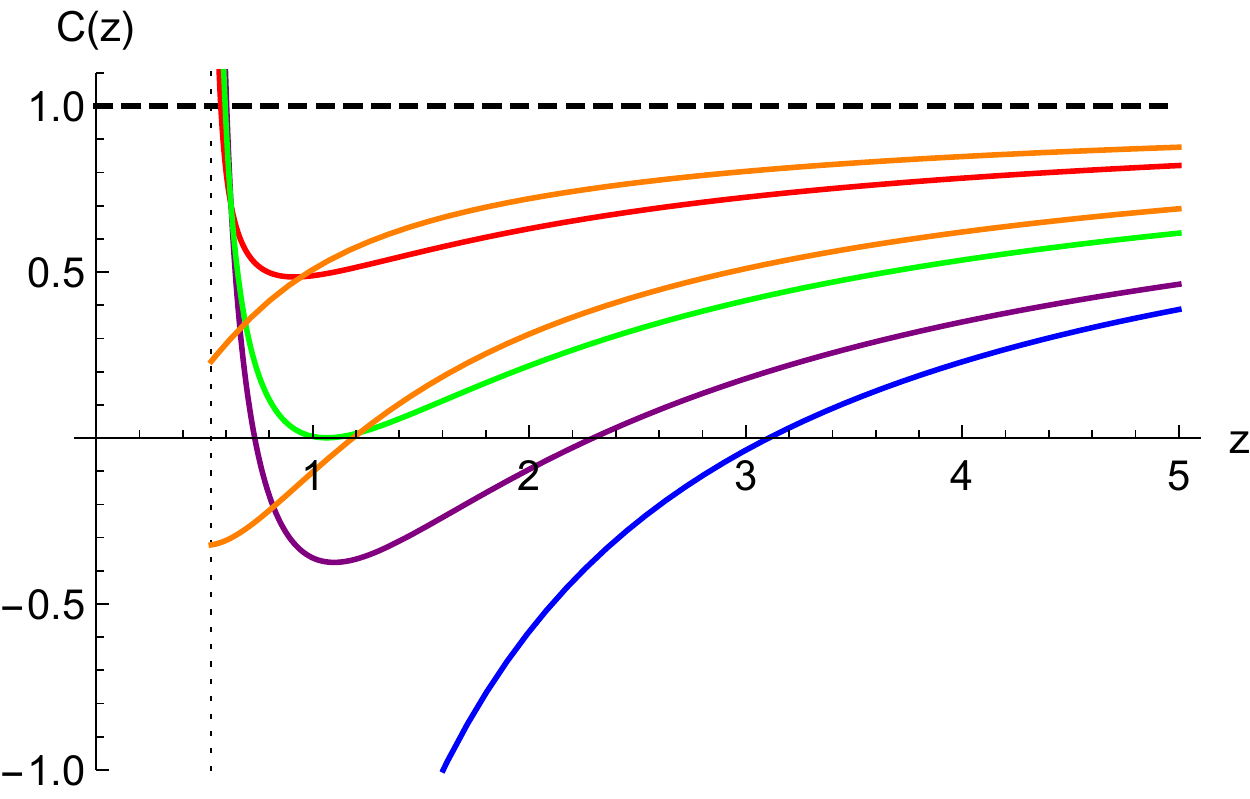}
	\caption{The metric function $C(z)$ in Eq.(\ref{eq:Final-ansatz-EiBI}) for $ l_\epsilon^2 =-1$ (for which $\delta_c \approx 3.260$) and the choices of $\{\delta_1=1,\delta_2=1/3\}$ (blue), $\{\delta_1=5,\delta_2=1/3\}$ (violet), $\{\delta_1=5,\delta_2 \approx 0.4675\}$ (green) and $\{\delta_1=5,\delta_2=1\}$ (red), corresponding to Schwarzschild-like black holes with a single horizon, and the three Reissner-Nordst\"om like configurations: black holes with two horizons, extreme black holes, and naked solutions, respectively. The two orange lines starting from a finite value of $C(z)$ at $z=z_c$ are Minkowski-like configurations ($\delta_1=\delta_c$) with a single horizon ($\delta_2=0.6$) or none ($\delta_2=1.5$). All solutions are asymptotically flat, $C(z) \underset{z \to \infty}{\approx} 1$. The vertical dotted line represents the wormhole throat $z=z_c$, to which all curves converge.}
	\label{fig:met2}
\end{figure}

\subsection{Properties of the solution: geodesic behaviour and regularity}

\begin{figure}[t!]
	\centering
	\includegraphics[width=0.40\textwidth]{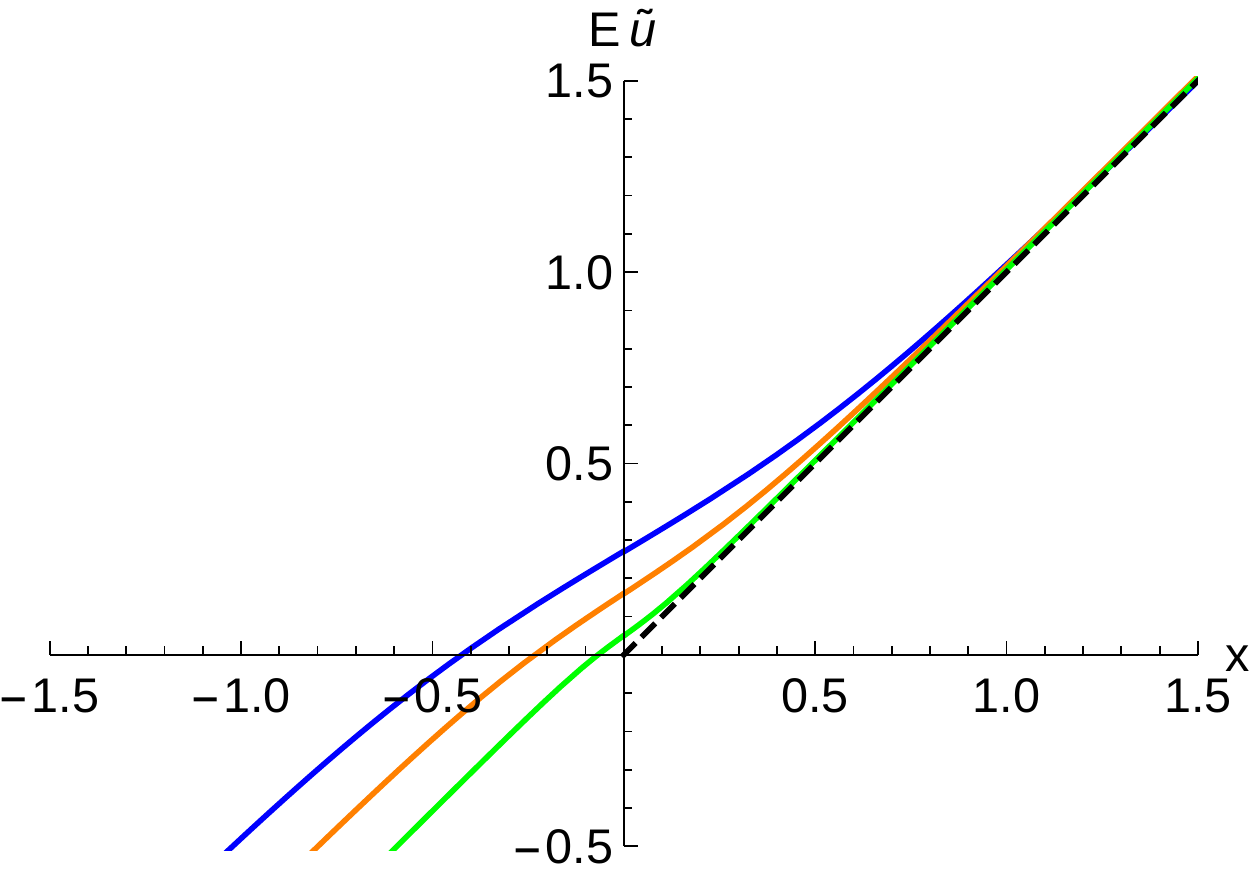}
	\caption{The affine parameter $ E \cdot \tilde{u}(x) $ versus the  radial coordinate $ x $ for  null radial geodesics. The green curve corresponds to $ \vert l_\epsilon^2 \vert=0.01$, the  blue curve to $ \vert l_\epsilon^2\vert=1 $, and the orange to $ \vert l_\epsilon^2 \vert=0.2 $. The black dashed line corresponds to the GR behaviour, where these geodesics end at $x=0$ and are therefore incomplete. At the wormhole throat the affine parameter obeys (\ref{eq:affinex}) and can be smoothly extended across $x=0$.}
		\label{fig:geo}
\end{figure}

For EiBI gravity with spherically symmetric solutions of the kind studied here, the geodesic equation (\ref{Eq:General_geo}) can be cast, taking into account the line element (\ref{eq:finalsolEiBI}), as
\begin{equation} \label{eq:geoBI}
\frac{1}{\Omega_+ ^2} \left(\frac{dx}{du}\right)^2= E^2 -V(x) \ ,
\end{equation}
with the same notation and conventions as in the $f(R)$ case above. Again, for null radial geodesics it is more useful to write this equation in terms of the radial function $z$. To this end, we take a derivative in Eq.(\ref{eq:xrBI}), which allows to cast (\ref{eq:geoBI}) in such a case as
\begin{equation} \label{eq:geonull}
\pm E d\tilde{u}=\frac{\Omega_{-}^{1/2}}{\Omega_{+}} \left(1+\frac{z\Omega_{-,z}}{2\Omega_{-}}\right)dz \ ,
\end{equation}
where again $\pm$ refer to ingoing/outgoing geodesics. It seems not possible to obtain a integration of this equation to find a closed expression for $\tilde{u}(z(x))$ everywhere, but we can resort to series expansions around the wormhole throat $z=z_c$. A glance at Eqs.(\ref{Omega_+_zc}) and (\ref{Omega_-_zc}) reveals that $\Omega_+$ is there just a constant that will have no impact in the behaviour of the solutions, while $\Omega_{-}$ contains the key factor in $(z-z_c)$. Thus, a little algebra allows to find the expansion of (\ref{eq:geonull}) at $z=z_c$ as
\begin{equation}
\pm E d\tilde{u} \approx \frac{\omega_-^{1/2}z_c}{2\omega_+} \frac{1}{\sqrt{z-z_c}} \ .
\end{equation}
This can be right away integrated as
\begin{equation} \label{eq:affinex}
\pm E (\tilde{u} - \tilde{u}_0) \approx \frac{\omega_-^{1/2}z_c}{\omega_+} \sqrt{z-z_c} \approx \frac{x}{\omega_{+}} + \mathcal{O}(x^2) \ ,
\end{equation}
where in the last equation we have made use of the fact that, using (\ref{eq:xrBI}) and (\ref{Omega_-_zc}), the radial function can be expanded in series of $x$ as
\begin{equation}
z \approx z_c +\frac{x^2}{z_c^2 \omega_{-} } + \mathcal{O}(x^4) \ .
\end{equation}
Since the domain of definition of the radial coordinate $x$ is the entire real line, nothing prevents the affine parameter in Eq.(\ref{eq:affinex}) to cross the wormhole throat and be indefinitely extended to the asymptotic infinity $x = - \infty$. This is shown in Fig. \ref{fig:geo}, where we numerically integrate the geodesic equation (\ref{eq:geonull}) in full range, showing that any such geodesic starting from a certain $\tilde{u}_0$ at $x=+\infty$ departs from the GR behaviour as the wormhole throat, $x=0$, is approached, and continues its path to another asymptotically flat region of space-time, $x = - \infty$.  Therefore, null radial geodesics are complete in this geometry.

For time-like geodesics and for null non-radial geodesics, one needs to analyze the behavior of the effective potential according to the expansion of the metric function at $z=z_c$ ($x=0$), as follows from Eq.(\ref{C-EiBI-zc}). Using the expansion (\ref{C-EiBI-zc}) this reads
\begin{equation} \label{eq:effwth}
V_{eff}\approx -\frac{a}{\vert x \vert} -b + \mathcal{O}(x) \ ,
\end{equation}
with the constants
\begin{eqnarray}
a&=& \frac{3 (1+2 \tau^2_c) (\delta_c-\delta_1)}{2\,\delta_2 \delta_c  \left(3+4 \,\tau^2_c\right)}\left(-k+\dfrac{L^2}{r_c^2 z\,_c^2} \right) , \\
b&=&\frac{3 \left(\delta_2-\delta_1 z_c^2\right)}{2\, \delta_2 (2+\sinh 2\,h(z_c))}\left(-k+\dfrac{L^2}{r_c^2 z\,_c^2} \right).
\end{eqnarray}
Indeed, likewise the structure of horizons, the fate of any such geodesic depends on the ratio $\delta_1/\delta_c$.

For Schwarzschild-like configurations, $\delta_1<\delta_c$, the potential (\ref{eq:effwth}) is infinitely attractive and, therefore, any such geodesic crossing the event horizon of these configurations will unavoidably get to the wormhole throat in finite affine time. At such a point the geodesic equation (\ref{eq:geoBI}) behaves as
\begin{equation}
\frac{d\tilde{u}}{dx} = \frac{ \vert x \vert^{1/2}}{\omega_{+}a^{1/2}} + \mathcal{O}(x^{3/2}) \to \tilde{u}(x)=\frac{2x \vert x \vert^{3/2}}{3\omega_{+}a^{1/2}} + \mathcal{O}(x^{5/2}) \ .
\end{equation}
As the coordinate $x$ extends over the whole real axis, it is clear that these geodesics can be naturally extended across $x=0$ and will be therefore complete for any values of the parameters of the model within the constraint $\delta_1<\delta_c$. It should be stressed that, despite the geodesically complete character of these space-times, any extended observer crossing the wormhole throat will find curvature divergences of size $K \sim 1/(z-z_c)^3$ there, which are much weaker than their GR counterparts, $ K \sim 1/r^8$. Therefore, one might wonder what would be the fate of any such observer undergoing arbitrarily large tidal forces as it crosses the wormhole throat. This question has been raised in other geodesically complete space-times in the literature via the effects of large tidal forces upon time-like observers modeled as a congruence of geodesics, and also with the scattering of waves off the wormhole, finding that no unavoidable physical pathologies should be present \cite{Olmo:2015dba}. A similar analysis would be needed for the solutions found here in order to guarantee their physical consistence, which nonetheless lies beyond the scope of this work.

For Reissner-Nordstr\"om-like configurations, $\delta_1>\delta_c$, the effective potential (\ref{eq:effwth}) flips sign and it is infinitely repulsive at $z=z_c$. Therefore, any of these geodesics will bounce at some $z>z_c$ and will continue its path within the $x>0$ (or $x<0$) region, which is the same behaviour as the one found in the Reissner-Nordstr\"om solution of GR.

Finally, for Minkowski-like configurations, $\delta_1=\delta_c$, the expansions (\ref{C-EiBI-zc}) and (\ref{grr-EiBI-zc}) are not valid, since one needs to replace first this value of $\delta_1$ in the line element (\ref{eq:finalsolEiBI}) before making the expansion around $z=z_c$, which yields the result
\begin{eqnarray}
g_{tt} &\approx&-\dfrac{3 (1+\frac{z_c^2 \delta_c}{\delta_2})}{2}\left(\dfrac{ \cosh 2\, h(z_c)}{1+2 \cosh 2\, h(z_c)} \right)  + \mathcal{O}(z-z_c)  \\
g_{rr} &\approx& \dfrac{3 }{2 \left(1+\frac{z_c^2 \delta_c}{\delta_2} \right)} \left(\dfrac{ \cosh 2\, h(z_c)}{1 + 2 \cosh (2 h(z_c)  )}\right)+ \mathcal{O}(z-z_c)
\end{eqnarray}
This implies that the effective potential takes the same form (\ref{eq:effwth}) with $a=0$, going to a constant  as $V_{eff}\approx -b + c(z_c)x^2$, where $c(z_c)>0$ is a constant with an involved dependence on $z_c$. Therefore, any particle with energy $E$ above the maximum of this potential will be able to get to the wormhole throat. At that point, its affine parameter will behave as
\begin{equation}
\tilde{\lambda}(x) \approx \frac{x}{\sqrt{b+E^2}} +\mathcal{O}(x^3) \ ,
\end{equation}
and therefore will find no impediment to continue its trip to the $x<0$ region. Moreover, as opposed to the Schwarzschild-like and Reissner-Nordstr\"om-like  configurations, in this case curvature scalars are all finite at the wormhole throat.

In summary, we have shown that all null and time-like geodesics in these geometries (in the branch $l_\epsilon^2<0$) are complete, no matter the values of mass and charge of the solutions or the value of the EH scale. The mechanism is, however, different from the $f(R)$ case, in that now the wormhole throat is accessible to different sets of geodesics, but all of them can be smoothly extended across the region $x=0$. Therefore, these geometries represent nonsingular space-times.

\section{Conclusion and discussion}  \label{sec:V}

In this work we have considered two families of gravitational theories extending GR, namely, quadratic $f(R)$ gravity and Eddington-inspired Born-Infeld gravity, both formulated in metric-affine spaces and coupled to Euler-Heisenberg electrodynamics. These two gravity theories have been chosen due to the different way the new dynamics is fed by the matter fields: in the $f(R)$ case the new effects in the gravitational sector are oblivious to anything but to the trace of the stress-energy tensor, while in the EiBI case they are sensible to its full content. The static, spherically symmetric solutions for both settings were found starting from the Einstein-like representation of the field equations. Such solutions suggested that only a branch of them, corresponding to a certain combination of the signs of the gravity and matter parameters, may hope to yield nonsingular solutions. Therefore, we focused on the characterization of such a branch according to the behaviour of the metric functions on the innermost region of the geometries, on the horizon structure, and on the completeness of geodesics.

The main conclusion of this analysis is that both settings do yield null and time-like geodesically complete space-times for all the spectrum of mass and charge of the corresponding solutions, provided that the aforementioned constraint on the signs of the parameters is met. While in both cases the singularity-regularization is possible thanks to the presence of a wormhole structure, the mechanisms for the completeness of geodesics differ. In the $f(R)$ gravity case, which has the same structure of horizons as in the Reissner-Nordstr\"om solution of GR, the central region is pushed to an infinite affine distance, so null radial geodesics would take an infinite time to get there, while for time-like geodesics or null non-radial geodesics the presence of an infinitely repulsive potential near the throat prevents them getting near it. Thus, only half of the wormhole (which may be covered by two horizons, a single extreme one, or be naked) is available for travel within the $x>0$ and $x<0$ regions.

As opposed to the $f(R)$ case, for EiBI gravity the throat can be reached in finite affine time by some sets of observers, depending on the ratio $\delta_1/\delta_c$,
which classifies the corresponding configurations in terms of horizons as Schwarzschild-like, Reissner-Norstr\"om-like, or Minkowski-like. If we focus on Schwarzschild-like configurations, which have a single event horizon, then the wormhole is a one-way structure, pushing out any observer departed from (say) $x>0$ and crossed the event horizon to the other asymptotic region in finite affine time. For Reissner-Nordstr\"om-like configurations, no matter their number of horizons, one finds instead that, like in their GR counterparts, any time-like observer could only get as close to the throat as it energy permits (given the existence of the infinite potential barrier), while null radial geodesics would only require a finite affine to get to the throat and cross it. Finally, Minkowski-like configurations (with a single horizon or none) have a finite maximum of its effective potential, thus allowing any observer whose energy is larger than it to cross the wormhole throat. Though curvature divergences generally appear at the throat (except in the Minkowski-like configurations, where curvature scalars are well behaved everywhere), the fact that they are much weaker than their GR counterparts, $\sim (z-z_c)^{-3}$, together with the lessons from previous research in the topic showing that extended observers are not necessarily destroyed in the transit through such regions \cite{Olmo:2016fuc}, raises questions on their true meaning when both the matter fields and the trajectories of idealized observers are well behaved.

The results obtained in this work further support the suitability of some metric-affine theories to get rid of space-time singularities in a variety of settings with conservative modifications of the GR framework. Moreover, these two basic mechanisms for such a singularity avoidance are shared by several other theories, and in agreement with the results of model-independent analysis in spherically symmetric space-times \cite{Carballo-Rubio:2019nel,Carballo-Rubio:2019fnb}. There are several challenges following these results, such as its compatibility with the semiclassical calculations of Hawking's radiation and black hole evaporation, the unsettled issue of topology change raised from the formation of wormholes, or to what extend these results can be sustained when moving to axially symmetric (rotating) scenarios. The latter is of special interest should any effect of metric-affine gravity  able to leak to the near-horizon scale, in order to address any of the opportunities offered by multimessenger astronomy. Work along these lines is currently underway.

\section*{Acknowledgments}

MG is funded by the predoctoral contract 2018-T1/TIC-10431. DRG is funded by the \emph{Atracci\'on de Talento Investigador} programme of the Comunidad de Madrid (Spain) No. 2018-T1/TIC-10431, and acknowledges support from the Funda\c{c}\~ao para a Ci\^encia e a Tecnologia (FCT, Portugal) research grants Nos.  PTDC/FIS-OUT/29048/2017 and PTDC/FIS-PAR/31938/2017,  the spanish projects FIS2014-57387-C3-1-P and FIS2017-84440-C2-1-P (MINECO/FEDER, EU), the project SEJI/2017/042 (Generalitat Valenciana), and PRONEX (FAPESQ-PB/CNPQ, Brazil). This article is based upon work from COST Actions CA15117 and CA18108, supported by COST (European Cooperation in Science and Technology).

\end{document}